\documentclass[12pt,preprint]{aastex}
\usepackage{times}
\usepackage{epsfig}

\shorttitle{The Emergence of the Thick Disk}
\shortauthors{Brook, Gibson, Martel, Kawata}

\begin{document}

\title{The Emergence of the Thick Disk in a CDM Universe II:\\ Colors and
Abundance Patterns.}

\author{Chris B. Brook,\altaffilmark{1} Brad K. Gibson,\altaffilmark{2,3}  Hugo Martel,\altaffilmark{1} \& Daisuke Kawata\altaffilmark{2}}

\altaffiltext{1}{D\'epartement de physique, de g\'enie physique et d'optique,
Universit\'e Laval, Qu\'ebec, QC, Canada  G1K 7P4}

\altaffiltext{2}{Centre for Astrophysics \& Supercomputing, 
        Swinburne University,
	Hawthorn, Victoria, 3122, Australia}

\altaffiltext{3}{School of Mathematical Sciences, Monash University, Clayton,
        Victoria, 3800, Australia}

\begin{abstract}
The recently emerging  conviction that thick disks are prevalent in disk galaxies, and their seemingly ubiquitous old ages, means that  the formation of the thick disk, perhaps more than any other component, holds the key to unravelling the evolution of the Milky Way, and indeed all disk galaxies. In Paper~I, we
proposed that the thick disk was formed in an epoch of gas rich mergers, at high redshift. This hypothesis was based on comparing N-body/SPH simulations to a \
variety of Galactic and extragalactic observations, including stellar kinematics, ages and chemical properties.
Here examine  our thick disk formation scenario in light of   the most recent observations of extragalactic thick disks. In agreement, our simulted thick disks are old and relatively metal rich, with V-I colors that 
do not vary significantly with distance from the plane. Further, we show that our proposal results in an enhancement 
of $\alpha$-elements in thick disk stars as compared with thin disk stars, 
consistent with observations of the relevant populations of the Milky Way.  
We also find that our scenario naturally leads to the formation of an old metal weak stellar halo population with high $\alpha$-element abundances. 
\end{abstract}

\keywords{galaxies: evolution ---
galaxies: formation --- galaxies: halos --- galaxies: structure ---
numerical methods}

\section{INTRODUCTION}

The emerging conviction that thick disks are prevalent in disk
galaxies offers rich possibilities of new insights into the origin and
evolution of  such galaxies.
Using the current thick disk as a fossil record, holding details of
the processes occurring at early stages of galaxy formation has
become one of the keys to shedding light on the origin of disk
galaxies.     
The thick disk of the Milky Way was discovered over twenty years ago
\citep{bernstein79,gr83}. Relative to the thin disk,
thick disk stars are kinematically warm, and lag in rotation by
$\sim20-40\,\rm{km\,s^{-1}}$ \citep{cb00}.
Studies have consistently found thick disk stars are old,  almost
exclusively older than 10 Gyrs (e.g.
\citealt{gw85,edvardssonetal93,furhmann98,gs04}). Yet thick disk
stars are relatively metal rich, with the peak of the metallicity
distribution of thick disk stars  at ${\rm[Fe/H]}\sim-0.6$ 
(e.g. \citealt{cb00}). 
 Thick and thin disk stars are chemically well separated (e.g. 
\citealt{furhmann98,prochaskaetal00,tautvaisieneetal01,
fbl03,mashonkinaetal03,reddyetal03,sp03,gs04,bensbyetal05},
hereafter BFLI), with thick disk stars having notably higher
$\alpha$-elements, with a ratio of $\alpha$/Fe for given Fe content
systematically higher than thin disk stars. 
There is at most only a very weak gradient in metallicity with height
above the plane in the thick disk \citep{gwj95,rbk01},
in fact BFLI found that abundance trends appear to be
invariant with vertical height.          

Thick disks have now been shown to be prevalent in disk galaxies
(Dalcanton \& Bernstein 2002, hereafter DB02). 
Surface photometry of a sample of 47 edge-on disk galaxies 
in DB02 also showed that extragalactic thick disks are red.
The next step in studying extragalactic thick disks was recently
taken by Mould (2005, hereafter M05), who resolved the stellar
population of a nearby sample of edge-on disk galaxies, to reveal
the properties as a function of height above the plane.  All four
galaxies of the M05 sample have thick disks composed of red stellar
populations.
The gradient of color with height, $\Delta({\rm V-I})/\Delta{\rm Z}$ 
is almost zero, or slightly positive. From the color magnitude
diagram (CMD) analysis, M05 concludes that the thick disks consist
of old and relatively metal rich populations, and their stellar
population is independent of height.
A study of the stellar content of NGC 55
by Davidge (2005, hereafter D05) similarly finds that stars 
associated with the thick disk are old (ages $\sim10\,{\rm Gyrs}$), with
the majority being in the metallicity range $-1.2<{\rm [Fe/H]}<-0.7$.
A further recent study by \cite{tgd05} also find evidence for thick disks in the three nearby spiral galaxies for which they analyse AGB and RGB stars.  
The results of M05, D05 and \cite{tgd05} are consistent with the conclusions 
of DB02, as well as with
observations of the Milky Way thick disk.

 A disk galaxy simulated (hereafter, simulated galaxies will be referred
to as sGALS to avoid confusion with real galaxies) using our
chemodynamical galaxy formation code, {\tt{GCD+}}, was shown to have a
thick disk component (\citealt{brooketal04b}, hereafter BKGF). 
This was evidenced by the velocity dispersion versus age relation for
stars around solar radii, which showed an abrupt increase in velocity
dispersion at lookback time of $\sim8\,{\rm Gyrs}$, in excellent agreement
with observation. 
By examining and comparing these simulations with observations, 
we proposed that thick disks form from gas which is accreted during a
chaotic period of hierarchical clustering of gas rich "building blocks"
at high redshift (BKGF). This formation scenario was shown to be
consistent with  observations of both the Galactic and extragalactic
thick disks.  

In this study, we examine four disk sGALS, and compare
their thick disk stellar populations to recent observations of
extragalactic thick disks, paying particular attention to the
observations in M05. Age and metallicity gradients with height above the
plane, as well as integrated V-I colors are derived. Comparison is
also made with the studies of extragalactic thick disks in DB02 and
D05. Further, we examine the abundance ratio of $\alpha$-elements 
with respect to iron in the stellar populations of the thin and 
thick disks, as well as halos of our sGALS.
These will provide further important
tests of our thick disk formation scenario, and tie in to current
studies on the formation of the stellar halo component of disk galaxies.

\section{THE CODE AND MODELS}

Our simulated galaxies (sGALS) are formed using our galactic
chemodynamical evolution code, {\tt GCD+}. This
 N-body/smoothed particle hydrodynamics (SPH) code
self-consistently follows the effects of gravity, gas dynamics, 
radiative cooling, and star formation. 
{\tt GCD+} also takes into account
metal enrichment and energy released by both Type~II (SNe~II) and
Type~Ia (SNe~Ia) supernovae, as well as the metal 
enrichment from intermediate mass stars. 
Full details of {\tt GCD+} can be found in 
\cite{kg03}. We apply the {\it{Adiabatic Feedback Model}} 
described in \citet{brooketal04a}. The model assumes that
the gas within the SPH smoothing kernel of SNe II
explosions is in adiabatic phase,  
in a fashion similar to a model presented in \citet{tc00}.
This allows us to form late-type sGALS which each have a dominant disk
component, and a less massive, relatively metal poor halo 
component \citep{brooketal04a}. 

We employ here a semi-cosmological version of {\tt GCD+}  based upon the
galaxy formation model of \citet{kg91}.  The initial condition is
an isolated sphere which consists of dark matter and gas
and whose total mass is $M_{\rm tot}$. 
This top-hat overdensity
has an amplitude, $\delta_i$, at initial redshift, $z_i$, which is
approximately related to the collapse redshift, $z_c$, by
\begin{equation}
z_c=0.36\delta_i(1+z_i)-1
\label{zc}
\end{equation}

\noindent (e.g.\citealt{pad93}). 
Small scale density
fluctuations are superimposed on each sphere, parameterized by
$\sigma_8$.  These perturbations are the seeds for local collapse and
subsequent star
formation. To incorporate the effects of longer wavelength
fluctuations, a solid-body rotation corresponding to a spin parameter,
$\lambda$, is imparted to the initial sphere.  

We assume an Einstein~de-Sitter CDM model with $\Omega_0=1$, 
$\lambda_0=0$, $H_0=50\,\rm{km\,s^{-1}Mpc^{-1}}$,  
baryon fraction, $\Omega_b=0.1$, and $\sigma_8=0.5$.
(but see \S~4 below). We also assume a star formation efficiency 
$c_*=0.05$ \citep{kg03}. We run four models of sGALS (sGAL1-4),
changing arbitrarily the parameters of $z_c$, $M_{\rm tot}$ and $\lambda$. 
The parameter values adopted are summarized in Table~1.
Values for $z_c$ are all well within the range of
distributions  for formation times of Milky Way sized dark matter halos
expected in a concordance $\Lambda$CDM cosmology (e.g. \citealt{power03}).
Along with these variations in parameters, different random seeds 
for initial density perturbations are
incorporated in the initial conditions of each sGAL, 
creating evolutionary diversity in our sample.
The random seeds are deliberately chosen 
to ensure no major merger occurs in late epochs ($z<$1).
These initial conditions, with large values of $\lambda$,  lead to morphology of late-type
disk galaxies for all four sGALS.
We employed 38911 dark matter and 38911
gas/star particles for each model, making the resolution comparable to 
other recent studies of disk galaxy formation.

\begin{deluxetable}{ccccc}
\tabletypesize{\footnotesize}
\tablecaption{Model Parameters and the age range, $t_{\rm thick}$,
for the definition of thick disk stars in our study.\label{modp}}
\tablewidth{0pt}
\tablehead{\colhead{Galaxy} & \colhead{$z_c$} & 
\colhead{$M_{\rm tot} ({\rm M}_\odot)$} & \colhead{$\lambda$} & 
\colhead{$t_{\rm thick}$}
}
\startdata
sGAL1 &  1.8  & 5$\times$10$^{11}$ & 0.0675 & 10.0-8.5 \\
sGAL2 &  1.9  & 5$\times$10$^{11}$ & 0.0675 & 10.4-8.5 \\
sGAL3 &  2.0  & 1$\times$10$^{12}$ & 0.0600 & 11.4-9.0 \\
sGAL4 &  2.2  & 5$\times$10$^{11}$ & 0.0675 & 11.0-9.0 \\
\enddata
\end{deluxetable}

\section{RESULTS}

\subsection{Vertical Stellar Population Gradient for Thick Disk Stars}

Figure~1 shows $I$-band images of each sGAL
at the present, displayed both face-on (X-Y plane, where
the disk rotation axis is set to be Z-axis)
and edge-on (X-Z plane, lower panels). 
The four sGALS are reasonably uniform in
their morphology, although sGAL4 at first glance appears to have a
more prominent bulge than the other three. Each has a dominant young,
metal rich thin disk. The stellar halo components are old and metal
poor. The surface density profiles of disk component for all four sGALS
are reasonably well fitted by exponentials.  

Evidence that our four sGALS have thick disk components is provided by
the velocity dispersion-age relation. Figure~\ref{sigW} shows
the velocity dispersion in the direction perpendicular to the disk
plane (Z direction) as a function of age for stars in the ``solar
neighborhood'' of our four sGALS. The solar
neighborhood in sGALS is defined as an annulus bounded 
by $6<R_{\rm XY}<10\,{\rm kpc}$, and $|{\rm Z}|<1\,{\rm kpc}$, 
where $R_{\rm XY}$ is the radius in the disk plane. 
The observed velocity dispersion-age relation for solar neighborhood stars
are also shown as triangles with error bars, as read from
Figure~3 of \citet{qg01}, who use the
data of \citet{edvardssonetal93}.  The observed velocity
dispersion is relatively constant for the past 
$\sim9\,{\rm Gyrs}$\footnote{The more recent study by \citet{nordstrometal04}
  finds that {\it thin} disk stars do show an increase of velocity dispersion
  with age, but this does not impact the analysis here.}, but shows an
abrupt increase at $\sim10\,\rm{Grys}$ ago. 
The older stars with high velocity dispersion are recognized as
thick disk stars. The velocity dispersion for 
the four sGALS show qualitatively very similar behavior, 
with plateaus interrupted by abrupt increases of velocity dispersion, 
at lookback times of between 8 and 10 Gyrs. 
The implication is that each of our sGALS also has an old thick disk
component.
In sGAL1 and sGAL2, these abrupt increases are more recent, at $\sim$8
Grys ago, compared with sGAL3 and sGAL4, whose jump in velocity
dispersion occurred closer to 10 Grys ago. The differences in the
lookback time of these increases in velocity dispersion in our sGALS is
partly related to the differences in collapse redshift, $z_c$, in
our initial conditions.


We use the timing of these increases in velocity dispersion of our
sGALS as a pointer to the thick disk formation era of each
simulation. In Figures~\ref{f3a}-\ref{f3d}, we plot four snapshots
at different time-steps in the evolution of
our sGALS during these thick disk formation epochs. 
In each sGAL, these epochs are defined by multiple mergers of gas rich
galactic building blocks. 
The first panel in each plot shows a number of "building
blocks",  and it is clear that these are gas rich.
As is detailed in BKGF, we confirmed that the gas mass to stellar mass ratio 
within these building blocks is high at this epoch.
A turbulent period of mergers then ensues. At this time,  
gas which carries significant 
angular momentum  ends up a disk structure.
However, the gaseous disk is kinematically heated up by these 
perturbations of merging
building-blocks, and has high-velocity dispersion.
Consequently the stars born in the disk at this period have
high-velocity dispersion, and are subsequently  identified 
as thick disk stars. 

This epoch is also characterized by rapid star formation.
Global star formation rate (SFR) as a function of lookback time for
the sGALS is plotted in Figure~\ref{sfr}.  
The thick disk formation epoch shown in Figures~\ref{f3a}-\ref{f3d}
corresponds closely to the peak of SFR.
The peak of SFR for sGAL1 and sGAL2 is later than that for sGAL3 and sGAL4,
corresponding to the earlier thick disk formation epoch in sGAL3 and sGAL4,
as reflected in Figure~\ref{sigW} and Figures~\ref{f3a}-\ref{f3d}.
By the end of the thick disk formation epoch, the thin disk component
starts to be built up (see BKGF for more details). 

Following BKGF, we identify disk stars which form in this epoch as
thick disk stars. Thus, we refer to disk star particles 
forming in the period highlighted in Figures~\ref{f3a}-\ref{f3d} as thick
disk stars. The age range for the definition of the thick disk is
summarized in Table~\ref{modp}. 
We arbitrarily decided this age range, based on
velocity dispersion (Fig.~\ref{sigW}), snapshots
(Figs.~\ref{f3a}-\ref{f3d}), and star formation rates (Fig.~\ref{sfr}).
We confirmed that the results shown in this paper are not
sensitive to  modest changes in these age definitions.
An additional criterion, applied to extract disk stars, is a  rotational 
 velocity of greater than $50{\rm km\,s^{-1}}$, which 
eradicates halo stars which form in  the  period
that the thick disk is built up.
%

We confirm that the properties of 
the thick disk stars identified are
similar to those observed in the thick disk of the Milky Way
(see BKGF for more details).
The rotation velocity of thick disk lags
that of thin disk stars by $20-30{\rm km\,s^{-1}}$.
We derive thick disk scale-heights of 1.3, 1.4, 1.1, and 1.4 kpc for
sGAL1-4 respectively, 
compared to the thin disk scale-heights of 0.52, 0.55, 0.6 and 0.45 kpc.
The thick disk stars are uniformly old, as expected by its definition.
We will demonstrate below that the thick stars in sGALS also have 
 characteristic metallicities observed in the Milky Way and 
external disk galaxies.

The key questions of the current study of sGALS are the 
variation in  the metallicities, age, and hence colors 
of thick disk stars  as a function of the distance from the 
disk plane, i.e. height ($|{\rm Z}|$),
as well as how their abundance ratios differ from later forming
thin disk stars. 
Figure~\ref{gradmet} displays the mean iron abundance, which 
we call metallicity hereafter, of thick disk stars,
as a function of $|$Z$|$.
Thick disk stars of sGAL1, sGAL2 and sGAL4 have metallicities
in the range of $\langle{\rm [Fe/H]}\rangle$ between $-0.5$ and $-0.6$, 
closely resembling the thick disk of the Milky Way,
while thick disk stars of sGAL3 have a significantly lower metallicity,
$\langle{\rm [Fe/H]}\rangle\sim-0.8$.
This may be related to the fact that the thick disk of sGAL4
is significantly less massive than the other sGALS.
However, to examine this, we need a statistically significant sample,
and thus we are now trying to increase the number of sGALS
(Renda et al. 2005).
The important thing is that
very little gradient is present in Figure~\ref{gradmet} for the thick
disk star particles of any sGAL.


Figure~\ref{gradage} demonstrates the age of the thick
disk star particles against $|$Z$|$.
Little trend of age with height can be said to exist. 
This is not trivial, because the age range for the definition
of the age of the thick disk is wide enough to have 
more significant age gradient.
It looks a little odd that sGAL4 thick disk stars are older on
average than those of sGAL3, seemingly contrary to the time of the
epoch given in Table~\ref{modp}. The is due to the sharper downturn in SFR
in sGAL4, as seen in Figure~\ref{sfr}, which means that a larger
fraction of thick disk stars in sGAL4 were born at the early 
time of the thick disk formation epoch, compared with sGAL3.

As a result of a lack of  significant gradient in neither 
age nor metallicity with
height, the thick disk star particles have no real color gradient.
Figure~\ref{VI} shows the mean $V-I$ color as a function of $|\rm{Z}|$.
Here we derive the mean $V-I$ color for the giant branch stars 
(RGB/AGB) of thick disk populations of the sGALS.
Padova isochrones \citep{girardietal00} are used to derive the
colors. 
We obtain the isochrone of the average age and metallicities
seen in Figures~\ref{gradmet} and \ref{gradage} with
linear interpolations are taken between the ages and metallicities
provided by the Padova tables.
Then the $V-I$ colors are derived by taking the initial mass function
(IMF) weighted mean of the isochrone within the range of 
$-2.7<{\rm I}<-9.7$,
which corresponds to  the luminosity range applied in M05.
The red colors of the giant branch stars reflects 
the age and metallicity of the thick disk populations, 
and there is no color gradient. These results are 
in good agreement with the colors, and lack of color gradient, as 
observed in M05.

Figure~\ref{BRRK} shows the $B-R$ and $R-K$, colors of our
thick disk stars. Here the colors are derived from the integration
of the luminosity of all thick disk stars which are still alive at the 
end of the simulations (redshift zero).
Again, Padova isochrones are used to derive the colors.  
We see that these colors are relatively constant with height, at
$B-R\sim1.4-1.5$, and $R-K\sim1.9-2.2$. These colors allow
direct comparison with Figures~3 and 6  of DB02. Away from the plane, DB02
find a relatively narrow range of colors, with $B-R\sim$ 1.0-1.4 and
$R-K\sim2.0-2.6$. In this study we do not examine sGALS with the same
mass range as the 47 galaxies studied in DB02, but rather examine only
${\rm L}_*$ sGALS. Nevertheless, thick disks of the sGALS  have
colors which can explain the stellar populations observed in DB02 to 
envelope edge-on disk 
galaxies,  which they interpret as thick disk components. 

\subsection{[$\alpha$/Fe] vs. [Fe/H] for Thin and Thick Disk and Halo Stars.}

We next compare the abundance ratios of $\alpha$-elements with
respect to iron for different sGAL components, thin disk, thick disk 
and stellar halo.
 In order to best compare with observations, we look at
star particles within the  ``solar neighborhood", as defined
above. Thin disk stars are defined as stars younger than 7 Gyr and
rotating faster than $50{\rm km\,s^{-1}}$.
The halo stars are represented by retrograde stars which formed before
the end of the thick disk formation epoch as defined in Table~1. Note
that the stars which have already formed in the 
building blocks which subsequently merge during thick disk formation
have a strong tendency toward ending up in the halo
(see \citealt{brooketal04a})\footnote{The tendency for accreted 
stars to be circularized
and hence end up as part of the disk components, as highlighted by
e.g.\citet{mezaetal05}, will only occur for stars accreted after a
disk has already formed. The fraction is expected  to be low in the
Milky Way, because such stars are expected to have  low metallicity, and 
the metallicity distribution function of solar neighbor stars
suggests little population of such metal poor stars.}. 
The $\alpha$-element abundances of the
three components are plotted in Figures~\ref{alpha1}-\ref{alpha4}.
Abundances of Oxygen and Magnesium are used to define our 
$\alpha$-elements.   
The spread of  the
distributions in the [Fe/H] axis is taken to include approximately 90\%
of stars within each component.  

It is evident that in all four sGALS, thick disk stars have higher
$\alpha$-element abundances than thin disk stars, even where they
overlap in metallicity, [Fe/H].  
These thick disk abundances are characteristic of
dominant enrichment from SNe~II. 
These results match very nicely with recent observations, 
(e.g. \citealt{furhmann98,prochaskaetal00,tautvaisieneetal01,
fbl03,mashonkinaetal03,reddyetal03,sp03,gs04}; BFLI).
 Uncertainty in our models must be noted here, regarding 
star formation prescriptions, nucleosynthesis yields, and particularly
SNe~Ia time-scales, which are still little understood theoretically. 
It is, however,  very encouraging that the overall
trends of the difference in abundance patterns between thin and
thick disk in the Milky Way can be reproduced with our
thick disk formation scenario.    

Halo stars in our sGALS also have high $\alpha$-element abundances. As
shown in \citet{brooketal04a}, stars in the gas rich building blocks
will preferentially accrete to the halo. Those formed in building
blocks accreted at high redshift will have been enriched primarily by
SNe~II. Halo stars tend to have slightly higher [$\alpha$/Fe]
than thick disk stars. This seems to be because SNe~II yields from
higher mass stars have higher [$\alpha$/Fe] \citep{ww95}.
The thin disk stars, which form in the later, more quiescent
period in the evolution of our sGALS, are clearly 
formed from material more enriched by SNe~Ia, which produces iron,
and have lower [$\alpha$/Fe].

\section{DEPENDENCE ON THE COSMOLOGICAL MODEL}


In this section, we address two potential limitations of our simulations:
First, the fact that we assumed an Einstein~de-Sitter model instead
of the prefered ``concordance model,'' and second, the fact that our
simulations deal with the collapse of an isolated
top hat perturbation.

\subsection{The Cosmological Model}

In our simulations, the background cosmological
model was an Einstein-de~Sitter model with
$\Omega_0=1$, $\lambda_0=0$, and $H_0=50\,\rm km\,s^{-1}Mpc^{-1}$.
However, observations now favor a flat, cosmological constant model 
($\Lambda$CDM) with
$\Omega_0=0.27$, $\lambda_0=0.73$, and $H_0=71\,\rm km\,s^{-1}Mpc^{-1}$
\citep{bennettetal03}.
While using a different model would have a major effect on simulations
of large-scale structure formation from Gaussian random noise initial conditions,
we can show that, in the context of the galaxy formation model we are
using, the effect is negligible.

The cosmological parameters at redshift $z$ can be expressed in terms
of their present values using
\begin{eqnarray}
\label{ozo}
\Omega(z)&=&{\Omega_0(1+z)^3\over(1-\lambda_0+\Omega_0z)(1+z)^2+\lambda_0}\,,\\
\lambda(z)&=&{\lambda_0\over(1-\lambda_0+\Omega_0z)(1+z)^2+\lambda_0}
\end{eqnarray}

\noindent (e.g. \citealt{mw90}).
Using the ``concordance'' values $\Omega_0=0.27$, $\lambda_0=0.73$,
we find $\Omega=0.99996$, $\lambda=0.00004$ at the initial redshift $z_i=40$.
At redshift $z=3$, by which time our simulations have ``turned around'' we have $\Omega=0.959$, $\lambda=0.041$. Afterward, the system becomes  significantly denser than the background
universe and the cosmology becomes irrelevant. Even as late as  $z_c=1.8$,
the smallest collapse redshift from our initial conditions, we have 
$\Omega=0.886$, $\lambda=0.114$. Hence, the model is very close to an
Einstein-de~Sitter model from the initial redshift up to the redshift of
collapse.  In particular, once the
galaxies have formed, the repulsive force caused by the cosmological constant
in totally negligible. 

We still need to compare the evolutionary timescales, which are
determined by the Hubble constant $H_0$.
The mean background density is given by $\bar\rho=\bar\rho_0(1+z)^3$.
Using $\bar\rho\propto\Omega H^2$, we get 
$H=H_0(\Omega_0/\Omega)^{1/2}(1+z)^{3/2}$. We eliminate $\Omega_0/\Omega$ 
using equation~(\ref{ozo}), and get
\begin{equation}
H(z)=H_0\left[(1-\lambda_0+\Omega_0z)(1+z)^2+\lambda_0\right]^{1/2}\,.
\end{equation}

\noindent This gives us the time-evolution of the Hubble parameter.
Using this equation, we evaluate $H$ at the
collapse redshifts, using both the Einstein~de-Sitter model
and the concordance $\Lambda$CDM model. 
The results are listed in Table~\ref{hubble}. The values of $H$ at
these redshifts are lower for the $\Lambda$CDM model, even though the
present value $H_0$ for that model is larger.

\begin{deluxetable}{cccc}
\tabletypesize{\footnotesize}
\tablecaption{Hubble parameter at collapse redshift (in units of 
$\rm km\,s^{-1}Mpc^{-1}$)\label{hubble}}
\tablewidth{0pt}
\tablehead{\colhead{Galaxy} & \colhead{$z_c$} & 
\colhead{$H(z_c,{\rm EdS})$} &
\colhead{$H(z_c,\Lambda{\rm CDM})$}
}
\startdata
sGAL1 &  1.8  & 234 & 183 \\
sGAL2 &  1.9  & 247 & 192 \\
sGAL3 &  2.0  & 260 & 201 \\
sGAL4 &  2.2  & 286 & 220 \\
\enddata
\end{deluxetable}

Consider first the redshift interval $[z_i,z_c]$.
As we argued above, the $\Lambda$CDM model is very close to the
Einstein-de~Sitter model in that range (that is, $\Omega\approx1$,
$\lambda\approx0$). However, the Hubble parameter has a different value,
and this implies a different evolution timescale.
In the redshift range $[z_i,z_c]$,
our Einstein-de~sitter model is {\it equivalent} to a $\Lambda$CDM
model in which the system would evolve too fast
because of the different
Hubble parameter. This turns out to have no
actual consequence. The relationship between initial and collapse
redshift (eq.~[\ref{zc}]) does not depend on any timescale. The collapse
redshift is entirely determined by the dimensionless parameters
$z_i$ and $\delta_i$. In our simulations, the background universe
evolves faster than in the $\Lambda$CDM model (by about 30\%),
but the top hat perturbation itself also evolves faster, such that
the collapse redshift will be the same in both models.

Consider now the interval $[z_c,0]$ between the
collapse redshift and the present. In this redshift interval, the system is
essentially isolated and the cosmology is irrelevant. All we need to do
is to compare the elapsed time between $z=z_c$ and $z=0$ to check that
the simulations terminated at a time that truly corresponds to the present.
For flat cosmological models
($\Omega_0+\lambda_0=1)$, the age of the universe at redshift $z$ is
given by
\begin{equation}
t={2\over3\lambda_0^{1/2}H_0}\arg\sinh
\left[\left({\lambda_0\over\Omega_0}\right)^{1/2}(1+z)^{-3/2}\right]\,.
\end{equation}

\noindent
Using this formula, we can compute the lookback time $t(0)-t(z)$ at
$z=z_c$ for our four simulated galaxies.
The results are shown in Table~\ref{lookback}. The differences between
the two models is only of order $2-3\%$, which is negligible.

\begin{deluxetable}{cccc}
\tabletypesize{\footnotesize}
\tablecaption{Lookback times (in units of Gyrs)\label{lookback}}
\tablewidth{0pt}
\tablehead{\colhead{Galaxy} & \colhead{$z_c$} & 
\colhead{lookback time (EdS)} &
\colhead{lookback time ($\Lambda$CDM)} 
}
\startdata
sGAL1 &  1.8  & 10.25 & $\phantom0$9.97 \\
sGAL2 &  1.9  & 10.40 & 10.16 \\
sGAL3 &  2.0  & 10.53 & 10.28 \\
sGAL4 &  2.2  & 10.76 & 10.62 \\
\enddata
\end{deluxetable}

Hence, we conclude that having considered an Einstein~de-Sitter model
instead of the concordance $\Lambda$CDM
model does not invalidate our simulations.

\subsection{Merger Histories}
The semi-cosmological model used in this study offers the advantage of allowing
several galaxies, which include detailed chemical enrichment histories,  to be studied within a hierarchical context. Here we do a larger survey of the merger history of halos using  N-body simulations in a fully cosmological concordance model. This will allow us to examine further the merging aspects of our proposed thick disk formation scenario with greater statistics in a less idealized cosmological context.

We perform N-body simulations of 
structure formation in a $\Lambda$CDM model with
$\Omega_0=0.27$, $\lambda_0=0.73$, and $H_0=71\,\rm km\,s^{-1}Mpc^{-1}$,
using a $\rm P^3M$ algorithm \citep{efstathiouetal85,martel91}.
In our initial run, we simulated a comoving cubic volume of size
$L_{\rm box}=100\,\rm Mpc$, using $128^3$ particles. The mass per particle
was $1.801\times10^{10}{\rm M}_\odot$. In this simulation, we identified
a region of size $\rm 30\,Mpc\times35\,Mpc\times28\,Mpc$
in which several Milky-Way-like halos formed, and we zoom-in by increasing
the resolution of the code by 2 in each direction in the region of interest
and rerunning the simulation. The simulation with zoom-in has a total of
2.5~million particles, and the mass per particle in the region of interest
is $2.251\times10^9{\rm M}_\odot$. This is sufficient to resolve the relatively large building blocks which we are interested in, in this study. We found in the region of interest 58 halos with masses in the range $(0.5-1.1)\times10^{12}{\rm M}_\odot$.

We have computed the merger trees for all 58 halos which are
Milky Way analogs. Two  trees are shown in Figure~\ref{trees}.
On each panel, the area of the circles are proportional to the mass
of the halos, and time increases from top to bottom. 
The black circle identifies a 
particular halo that formed by a ``major'' merger or a series of ``significant'' mergers. Halo~1 formed by the merger of 5 halos
of comparable masses occurs between 9.3 and 7.7 Gyrs ago (resembling sGAL1). 
For halo~2, several mergers take place between 11.0 and 10.3 Gyrs ago (and may be compared with sGAL3); this involves 4 halos of significant mass. Another  merger occurs between 9.3 and 7.7 Gyrs ago, and a smaller one between 7.7 and 5.0 Gyrs ago. In this study, we are interested in merger events
which are not necessarily ``major mergers'', in which halos of comparable mass are involved, but rather mergers which we may call ``significant'', in that they add a significant amount of mass to the forming central halo.
In order to quantify this, we define ``significant'' in this section as a ``building block'' with mass $\ge 10^{10}\rm{M}_\odot$ which adds at least $4\%$ to the  mass of the forming halo. Such a merger event is able to  
disturb the  disk structure which is forming (\citealt{qhf93}). We plot the average number of such ``building blocks'' versus lookback time in Figure~\ref{buildingblocks}. In order to qualify, the ``building block'' must be merged to the central halo\footnote{or largest halo; at early times there are often several similar sized halos, like in halo~1 of Figure~\ref{trees}}  by the next time interval, so that Figure~\ref{buildingblocks} summarizes the merger history of the final halo.  The number of ``building blocks'' at each time interval includes the central halo, so a value of one on the Y-axis corresponds to there being only a central galaxy. We see that, on average, halos of mass comparable to the Milky Way only have a few ``significant'' mergers, and that  most occur between $\sim8$ and $\sim$11 Gyrs ago. We include all 58 final halos of mass
$(0.5-1.1)\times10^{12}{\rm M}_\odot$ in this plot. Several, ($\sim 15$), halos have major mergers at late epochs (less than 5 Gyrs ago), and are thus probably not good candidates for hosting late type galaxies. Omitting these galaxies from our study would extenuate our findings in this section that the bulk of the ``significant'' merging events of Milky Way type dark halos occur at high redshift.

\section{DISCUSSION}

Several scenarios have been put forward to explain the origin of the
thick disk of the Milky Way, which has become established as a separate
component from the thin disk . A review of these was made in BKGF and
will not be repeated here. Readers are also referred to reviews by
\citet{gwk89} and \citet{wyse04}. The importance of thick disk
formation has seemingly been enhanced by recent observations suggesting
that thick disks are prevalent in spiral galaxies, with their seemingly
ubiquitous old age implying that their formation may be a key in
unraveling the formation of such galaxies. In BKGF,
we proposed 
a  scenario in which the bulk of thick disk stars forms in a period of 
multiple  merging of gas-rich building blocks at high redshift,
prior to  thin disk stars forming. 
As this is very much a follow up paper to BKGF, we
will not repeat our analysis of the various thick disk scenarios, and
their consistency with current observations. Suffice to say that we
believe  our scenario to be consistent with both Galactic and
extragalactic observations. In this paper we have analyzed our thick
disk formation scenario in light of new results in M05, as well as
explicitly examining $\alpha$-element abundance signatures.

 By resolving individual stars in four local edge-on disk galaxies with the
{\it Hubble Space Telescope}, M05 confirmed  that a) thick disks are common
features of spiral galaxies, b) thick disk stars are old, c)
relatively metal rich and, d) have no significant vertical metallicity
gradient.  These observations do not support thick disks
forming from shredded stars from accreting satellites (because
they should be metal poor), or slow heating of the thin disk
(a color gradient with height, and young thick disk stars would be expected). 
Lack of  detection of a vertical metallicity
gradient also puts a strong constraint on a scenario which
involves {\it slow} dissipative collapse.
We have shown in Figures~\ref{gradmet} and \ref{gradage} that our thick
disk formation scenario is consistent with these observations. 
Especially, there is no vertical trend in the metallicity or age,
and thus color. This is because in our thick disk formation 
scenario the thick disk forms from well-enriched gas disk in situ,
in a relatively short period. Astration during and after
the mergers, as well as the ongoing dissipative infall of gas (see e.g.
\citealt{muralietal02}) eliminates difference of metallicity of thick disk
forming material,  which leads to a relatively homogeneous metal rich
and old stellar population, independent of vertical height.
The M05 result is supported by other recent studies, D05 and  \cite{tgd05}.


Our thick disk stars have a combination of ages and metallicities
which are able to  explain the colors
which DB02  find in the diffuse envelope
components they observe in  their sample of 47 edge-on disk galaxies, 
as shown in Figure~\ref{BRRK}.
They believe that these envelopes are best explained by the prevalence
of thick disks in spiral galaxies.  

A study by \citet{sddj05} of 6 nearby edge-on late type 
galaxies with resolved
stellar populations also finds evidence for a thick disk or halo
population, but their preliminary analysis of red giant branch colors
indicate that this stellar envelope is more metal poor than those found
by M05 and D05, with metallicities ranging between
$-1.2<{\rm[Fe/H]}<-1.7$. This might suggest a diversity of the 
stellar population of the thick disks among different disk galaxies.
Our four sGALs do not have such low metallicity. However, more
statistical sample of sGALS with different masses and merger histories
might explain such observations.

The building blocks which merge at high redshift in our thick disk
formation scenario have primarily been enriched by SNe~II. We have shown
in Figures~\ref{alpha1}-\ref{alpha4} that our thick disk formation scenario naturally
explains the high $\alpha$ enhanced abundance pattern observed in thick disk
stars compared to thin disk stars (e.g.\citealt{reddyetal03,gs04}; BFLI). 
This is because in our scenario thick disk stars
form earlier than thin-disk stars, which
have more chance to be enriched more by SNe~Ia.
The other important fact affecting enhanced $\alpha$-element abundances  is 
 high star formation rates  during
the merger epoch in which the thick
disks form in our sGALS. Evidence that the thick disk has had a more
intense star formation history than the thin disk was found by BFLI,
and \citet{mashonkinaetal03}.

In order to simulate several sGALS with high resolution
required for this study, use was made of simplified initial
conditions. To place our study on a firmer cosmological background, we have
used N-body simulations employing the concordance $\Lambda$CDM values
to show that Milky Way sized  dark halos evolve in a manner which
supports our thick disk formation scenario. Most of the mass of such halos are
accreted from the merging of a few intermediate sized
($> 10^{10}{\rm M}_\odot$) building blocks, with a period of peak
merging activity of such halos at lookback time of $\sim8-11{\rm Gyrs}$.
This is consistent with the study of \citet{zb03},
who find that dark halos of mass $\sim1.4\times10^{12}{\rm M}_\odot$
have a characteristic phase of mergers/disruption of subhalos of mass
$\sim1-10\times10^{10}{\rm M}_\odot$  at a lookback time of $\sim10\,{\rm Gyrs}$. In
fact, \citet{zb03} remark on the proximity in time of this
merger phase with the ages of Milky Way and extragalactic thick
disks. The timescale of our merger period is also consistent with distributions  for formation times of Milky Way sized dark matter halos
expected in high resolution simulations of  concordance $\Lambda$CDM cosmology (e.g. \citealt{power03}).

Our thick disk formation scenario is nicely coupled with
the formation of an old, metal weak stellar halo
population with $\alpha$-element enhancement.
In \citet{brooketal04a}, accretion of stars to the halo in gas rich
merger events was highlighted as necessary in creating low mass, low
metallicity stellar halos in sGALS. Hence, many such halos stars
formed in building blocks prior to the multiple merging epoch.
The relatively low masses (recall the well established mass-metallicity 
relation for galaxies), and short times afforded for star formation prior 
to merging, ensure that these stars are metal poor.  
This can explain the oldest, and most metal poor halo stars.
Figures~\ref{alpha1}-\ref{alpha4} show that the halo stars will also
have $\alpha$-element enhancement. This halo star formation
scenario was also highlighted in a recent semi-analytical study by
\citet{robertsonetal05}, who showed that  building blocks which merge
at high redshift are likely to be a few relatively large
($\sim5\times10^{10}$) dwarf irregular analogs, with 
$\alpha$-element enhancement.       

We feel it necessary to re-iterate  the comment made in BKGF that late,
direct accretion of stars after a disk has formed, in the manner
described in \citet{qhf93} can also play a role in 
contributing to the thick disk
population (e.g. \citealt{bkg03,yannyetal03,martinetal04,nhf04,mezaetal05}). 
Such accretion has been shown to
play a role in formation of the stellar halo 
(e.g. \citealt{igi94,helmietal99,bkg03}).   

One recent study needs to be mentioned here. \citet{yd05}
present evidence for the existence of a counter-rotating thick
disk in FGC 227.  Existence of counter-rotating thick disks would rule out
models in which thick disks forms purely by monolithic collapse or
from heating of a previous thin disk, and strongly favors
accretion/merger models. The
initial conditions of our models, in which a solid body rotation is
imparted on our initial sphere, means that we cannot directly test for
the existence of counter-rotating thick disks in the sGALS presented
in this paper. All building blocks of significant mass which merge 
during thick disk formation epoch in our four sGALS have prograde rotation.  
However, it would not be surprising that in a multiple
merger period there would be building blocks which  have
counter-rotation. Hence,  our thick disk formation scenario remains  
a good candidate for explaining such observations. 
However, more statistical observations and simulations based
on full-cosmological context are required to test our scenario.

\acknowledgments
Firsly we thank Jeremy Mould whose advanced provision of results, and insightful correspondence, provided the impetus for this study. We also thank Peter Yoachim and Julianne Dalcanton for providing  results of their study on extragalactic thick disk ahead of publication. Simulations were performed on Swinburne University Supercomputer, those of the Australian Partnership for Advanced Computing, and on ``Purplehaze,'' the Supercomputer facility at Universit\'{e} Laval. CB and HM thank the Canada Research Chair program and NSERC for support. BKG and DK acknowledge the financial support of the Australian Research Council through its Discovery Project program.

\begin{figure*}
\epsfxsize=15.1cm \plotone{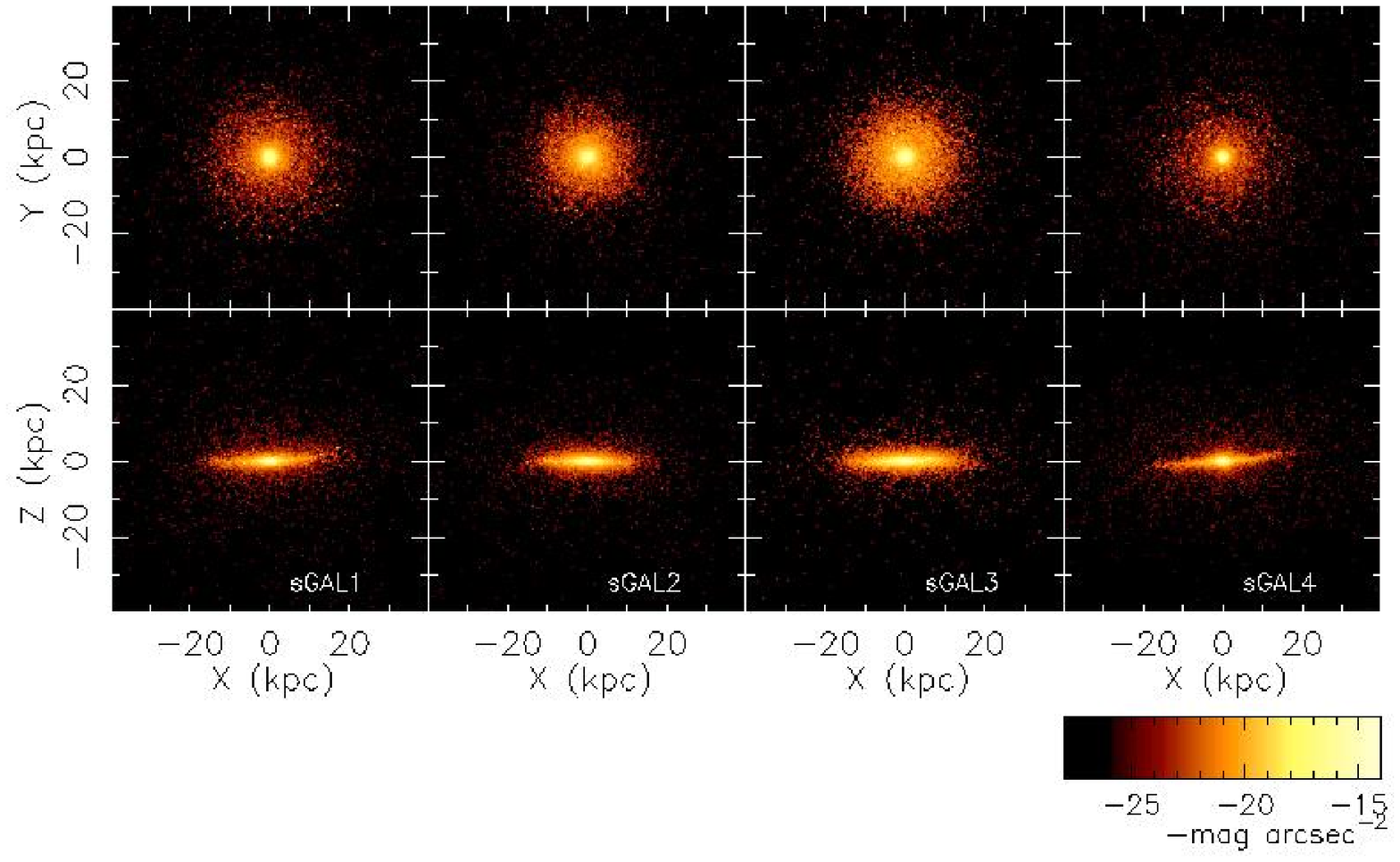}
\caption{I-band image for our 4 sGALS at $z=0$, shown in edge-on
 (X-Z plane, upper panels) and face-on (X-Y, lower) projections. 
The sGALS are dominated by young, metal rich thin disks. 
\label{fig1}}
\end{figure*}

\begin{figure}
\epsfxsize=140.mm \plotone{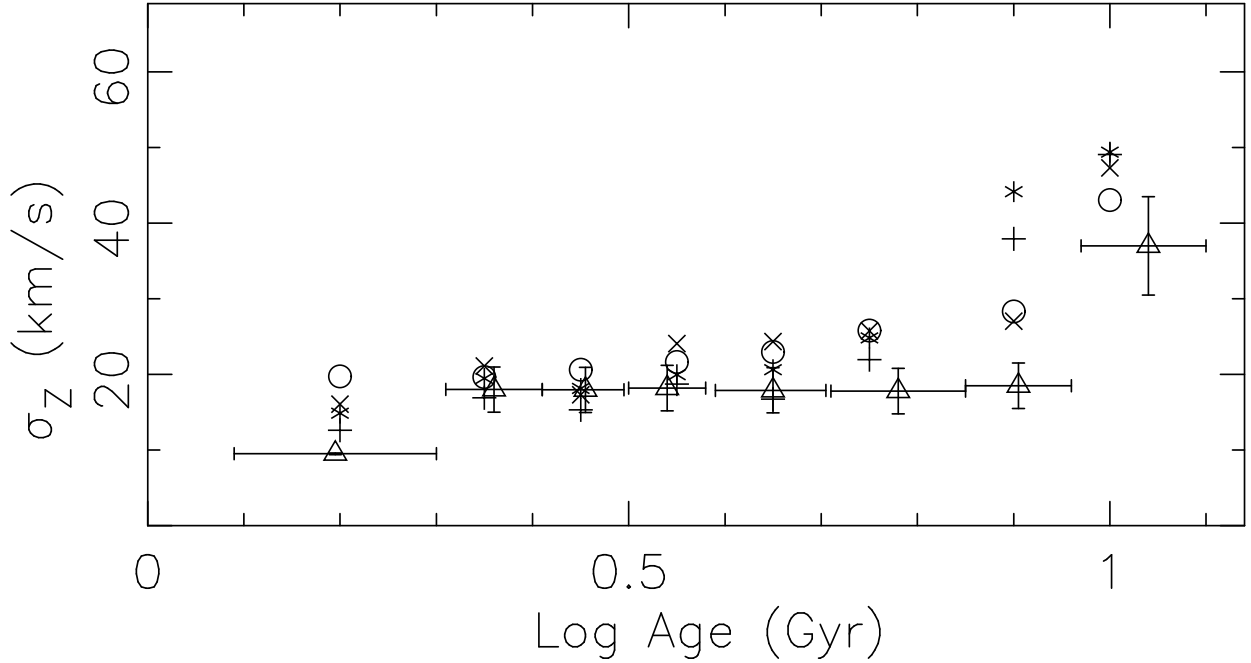}
\caption{ Velocity dispersion for the direction of 
perpendicular to the plane (Z direction) 
as a function of the age for four sGALS: sGAL1/2/3/4 is designated by symbols
$+$/$\ast$/$\circ$/$\times$, respectively.
 The velocity dispersion is measured around star particles
within $6<{\rm R_{XY}}<10$ kpc and $|$Z$|<$1 kpc.
Also shown by triangles with error bars are 
observations of solar neighborhood stars, as derived by
Quillen \& Garnett (2001, Fig.~3) from the sample of \citet{edvardssonetal93}.
In each of our sGALS, as well as the observations, an
abrupt increase in the velocity-dispersion is apparent at
lookback times $>8$ Gyrs. This abrupt increase is a signature of the
thick disk. sGAL1 and sGAL2 show an abrupt increase at lookback time
$\sim$8 Grys, while the upturn in velocity dispersion in sGAL3 and
sGAL4 is at an earlier time, closer to the lookback time of $\sim$10
Grys observed for the Milky Way.
}
\label{sigW}
\end{figure}

\begin{figure*}
\epsfxsize=140.mm \plotone{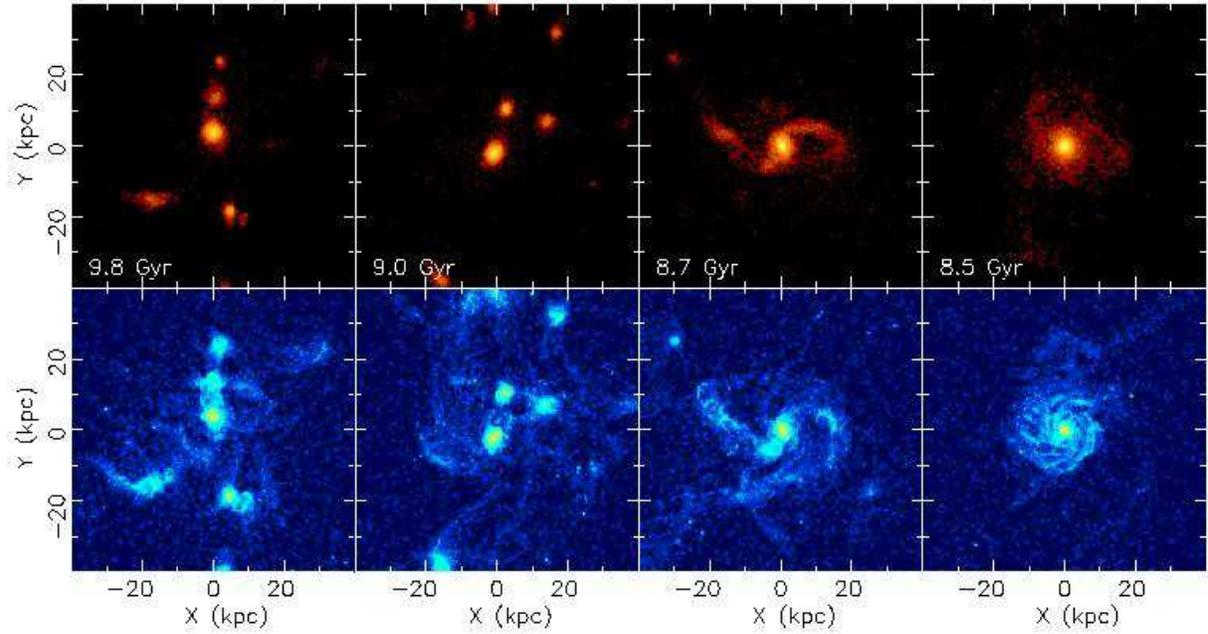}
\caption{Density plots of stars (upper panels) and gas (lower panels)
 of sGAL1 during the epoch of thick disk formation, shown face on (X-Y
 plane). Four time-steps are shown, at lookback times ranging from
 9.8-8.5 Gyrs. This epoch corresponds to the time at which the
 velocity dispersion-age relation shows a sharp increase, indicating
 the presence of a thick disk (Fig.~2), and is characterized by gas
 rich merger events. At the beginning of this sequence, several gas
 rich galactic "building blocks" exist, while at the end of this
 epoch, a central sGAL has formed. }
\label{f3a}
\end{figure*}
\begin{figure*}
\epsfxsize=140.mm \plotone{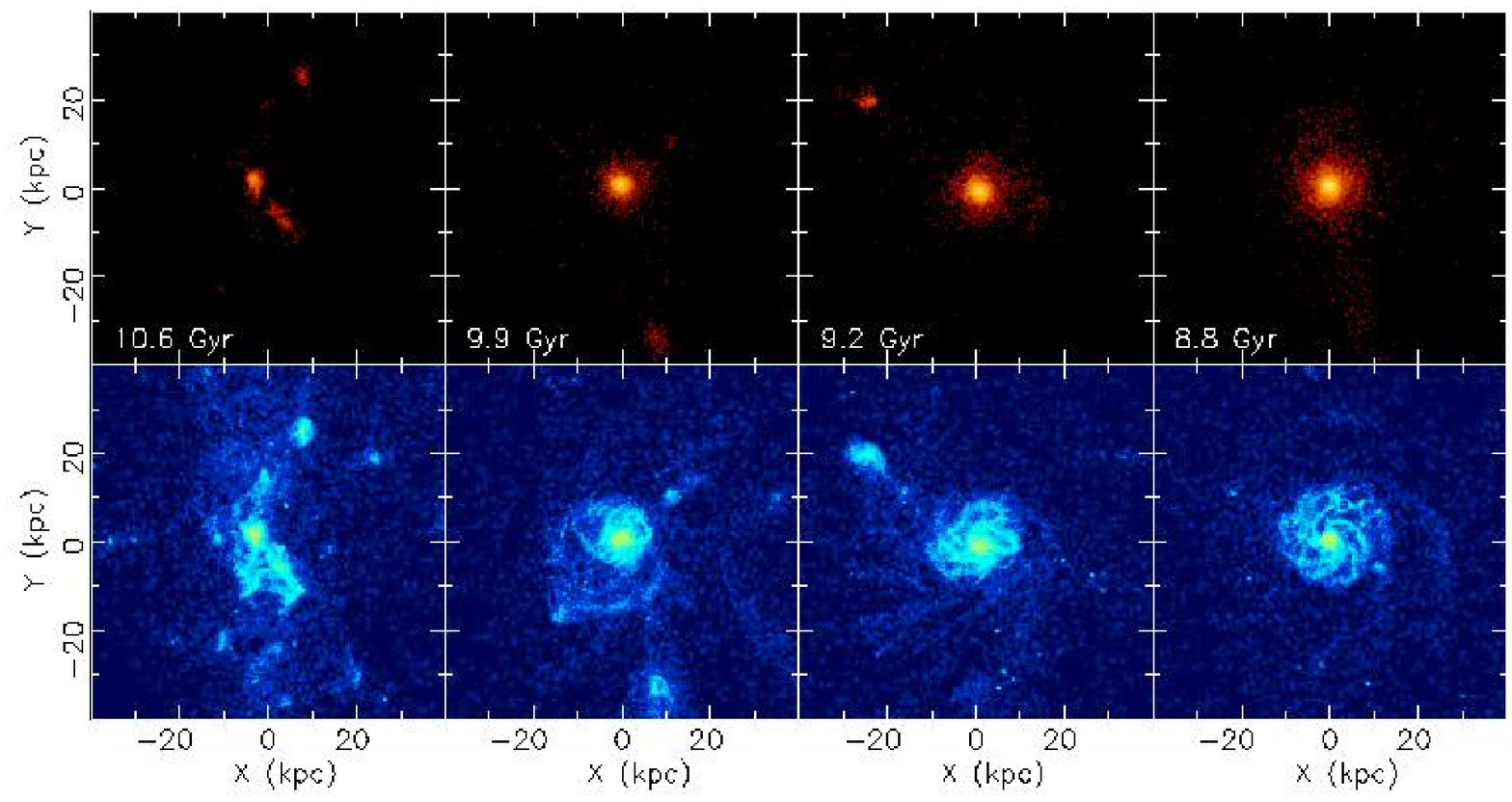}
\caption{Same as in Figure~\ref{f3a}, but for sGAL2. }
\label{f3b}
\end{figure*}

\begin{figure*}
\epsfxsize=140.mm \plotone{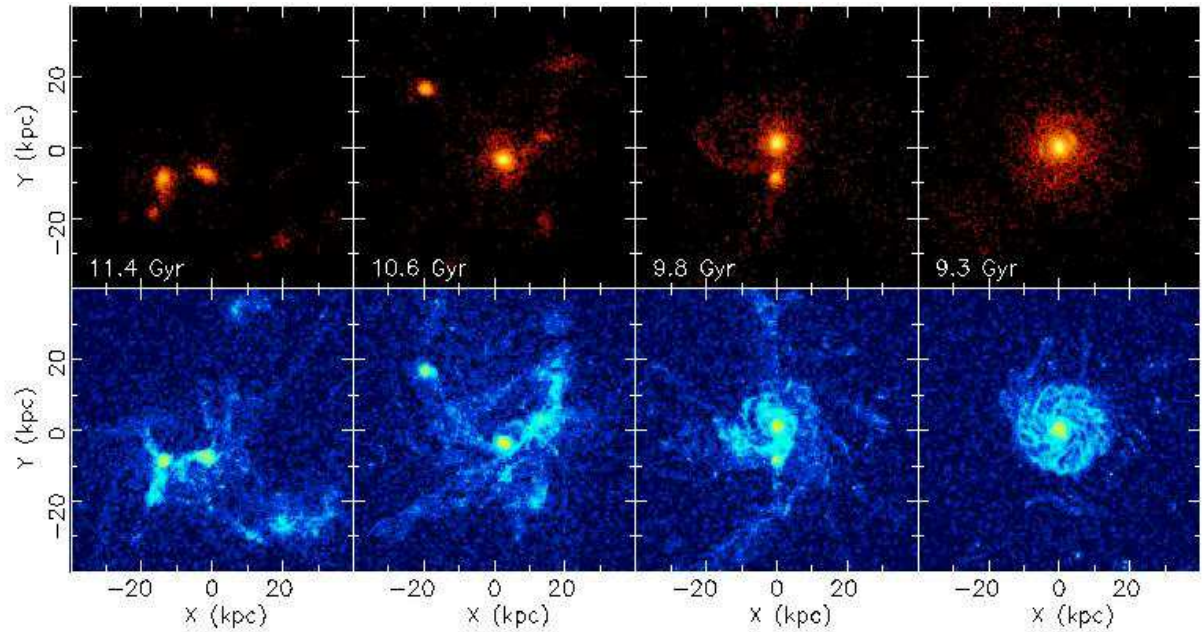}
\caption{Same as in Figure~\ref{f3a}, but for sGAL3. We see that the epoch of gas 
rich mergers which we associate with thick disk formation is earlier
 in this sGAL than in sGAL1 and sGAL2, consistent with the older
 lookback time of the abrupt increase in velocity dispersion seen in
 Figure~\ref{sigW}, 
as well as the time of the peak of the star formation rates
 (Fig.~\ref{sfr}). }
\label{f3c}
\end{figure*}

\begin{figure*}
\epsfxsize=140.mm \plotone{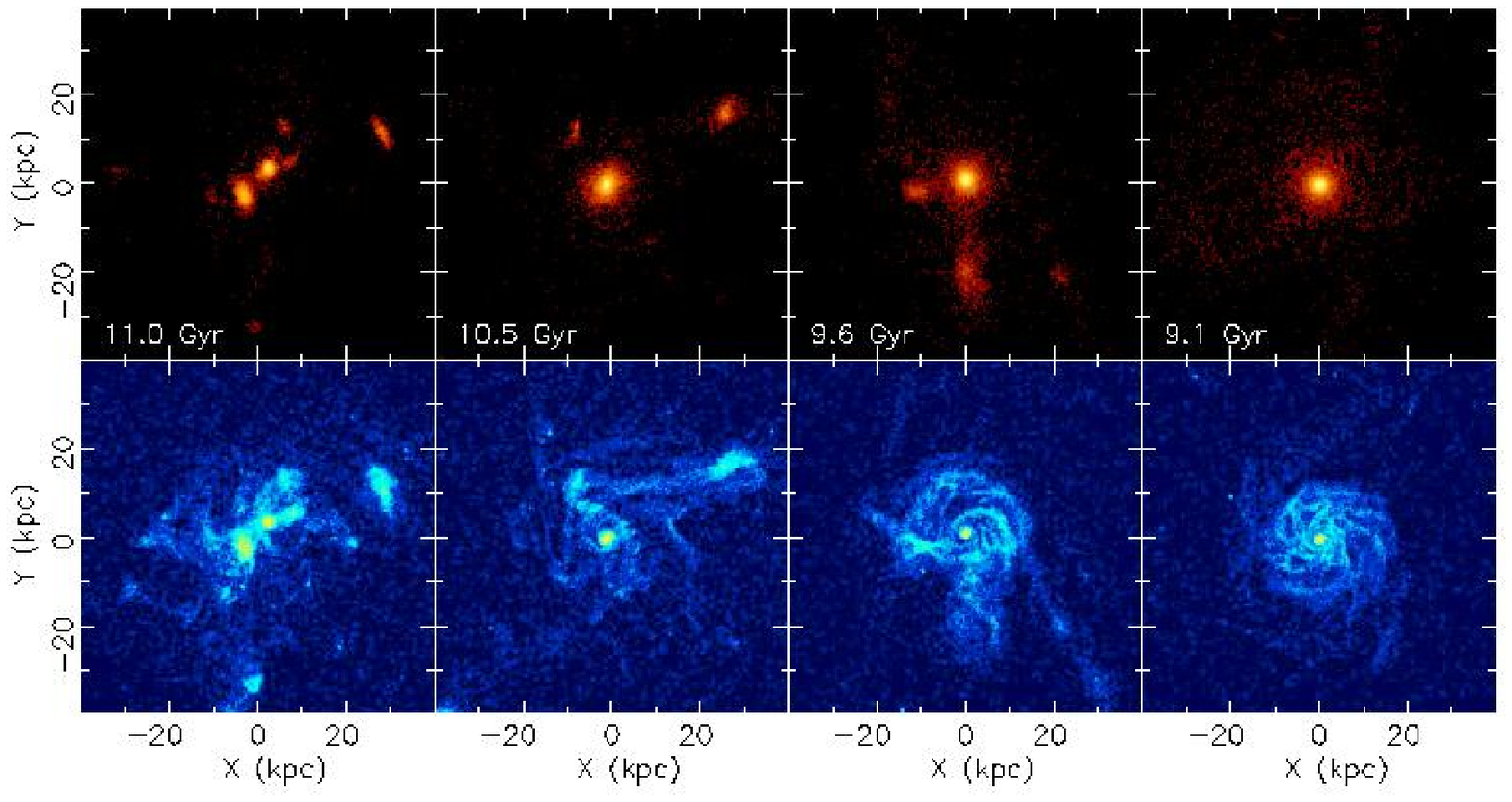}
\caption{Same as in Figure~\ref{f3a}, but for sGAL4.
}
\label{f3d}
\end{figure*}

\begin{figure}
\epsfxsize=140.mm \plotone{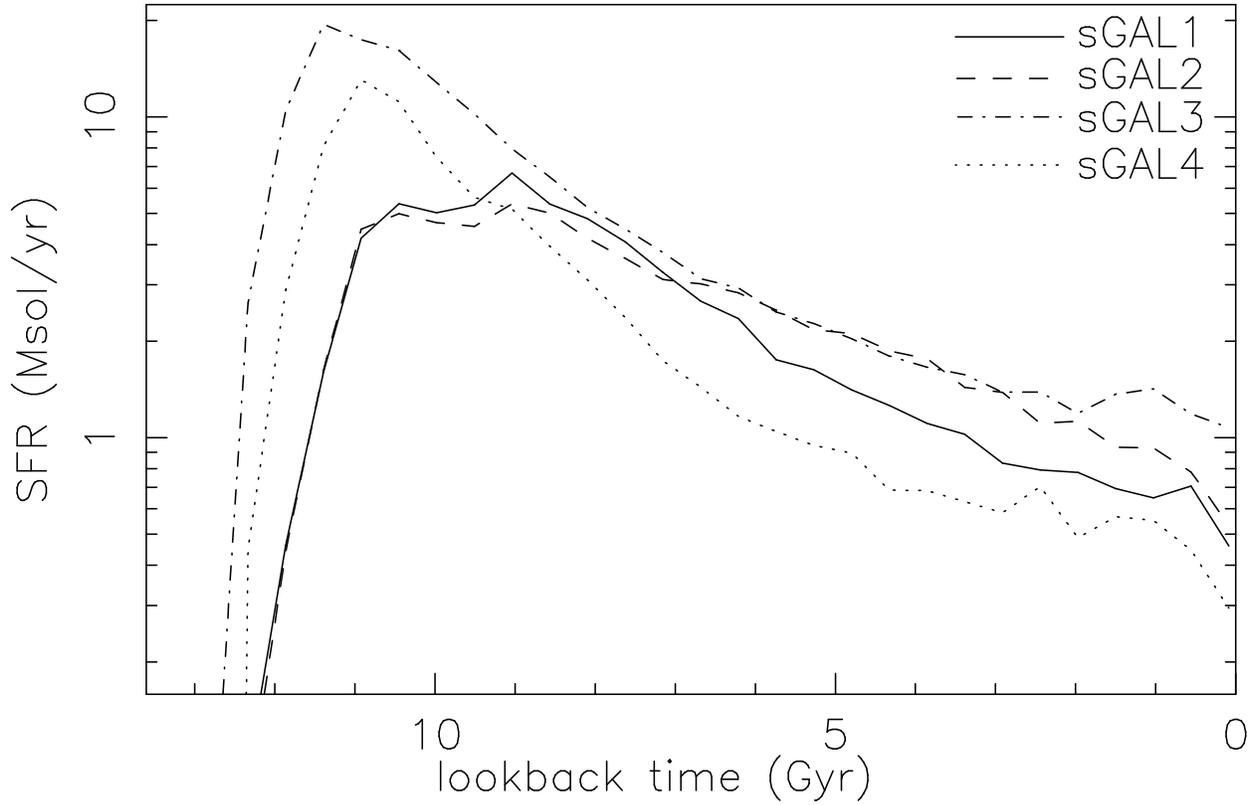}
\caption{Global star formation rate (SFR) as a function of lookback
time for sGALS. The solid, dashed, dot-dashed, and dotted
lines indicate the SFR of sGAL1, sGAL2, sGAL3, and sGAL4, respectively.
The peaks of the SFR correspond to the epochs of gas rich merging
which we associate with thick disk formation period.
Note that sGAL1 and sGAL2 have the peaks of the SFR
at later a epoch than sGAL3 and sGAL4, corresponding to the later
occurrence of the gas rich merging epoch in these sGALS, 
as shown in Figures.~\ref{f3a}-\ref{f3d}. }
\label{sfr}
\end{figure}

\begin{figure}
\epsfxsize=140.mm \plotone{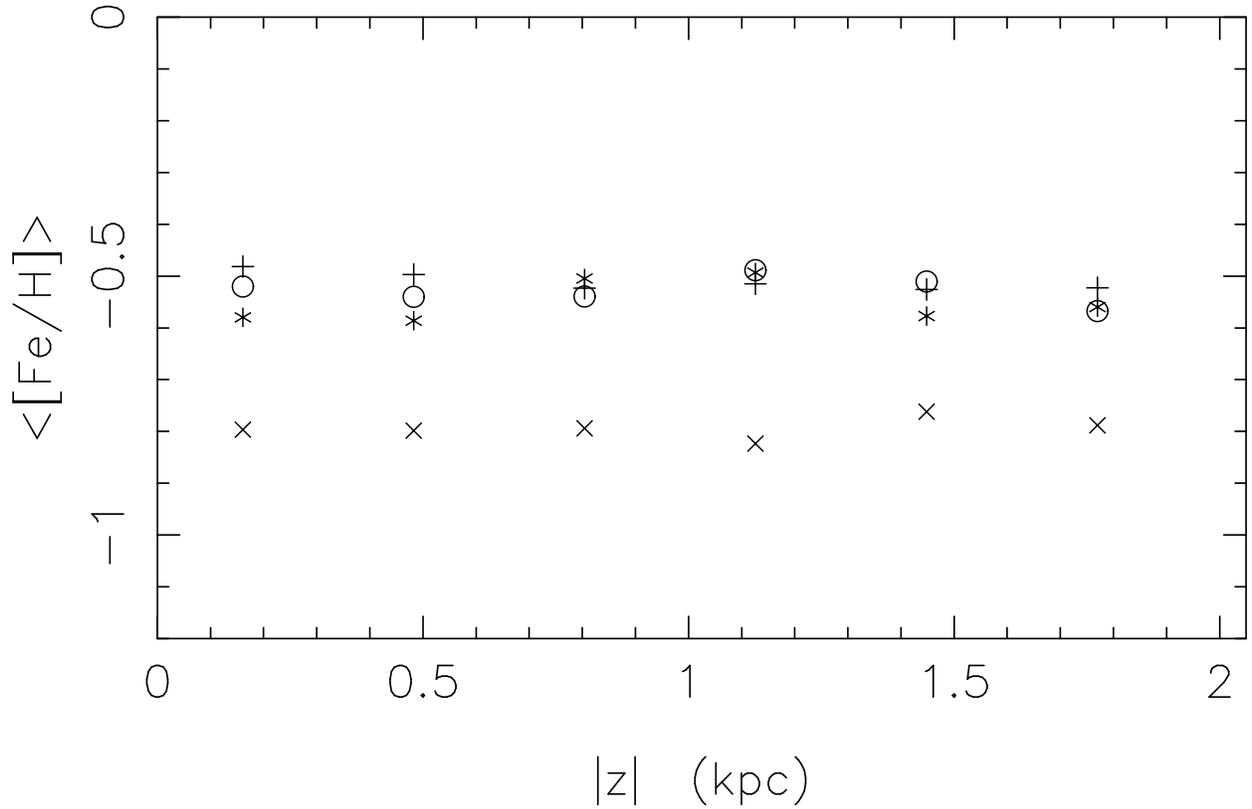}
\caption{Mean iron abundance ($\langle$[Fe/H]$\rangle$) 
as a function of the distance from the disk plane, i.e. height ($|$Z$|$)
for thick disk stars. The same symbols as Figure~\ref{sigW} are used.
Thick disk stars of sGAL1, sGAL2 and sGAL3 have
$\langle$[Fe/H]$\rangle$ between $-0.5$ and $-0.6$,
while thick disk stars in sGAL4 have a lower metallicity,
$\langle$[Fe/H]$\rangle\sim-0.8$. 
Very little gradient is apparent for the thick
disk star particles of all four sGALS.
}
\label{gradmet}
\end{figure}
\begin{figure}
\epsfxsize=140.mm \plotone{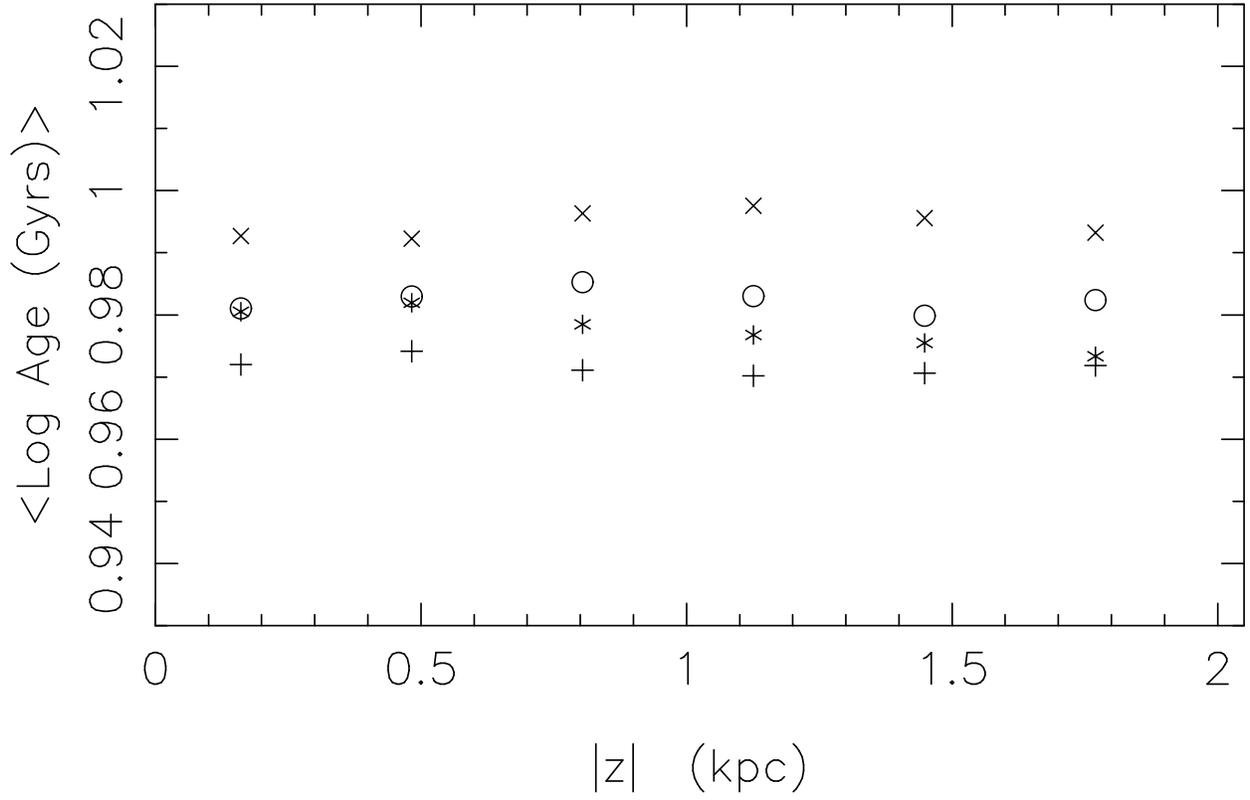}
\caption{Mean of logarithm of age (Gyrs) as a function of
$|$Z$|$ for thick disk stars. The same symbols as Figure~\ref{sigW}
are adopted. Thick disk stars in each galaxy are old, with little or no
variation with height.}
\label{gradage}
\end{figure}

\begin{figure}
\epsfxsize=140.mm \plotone{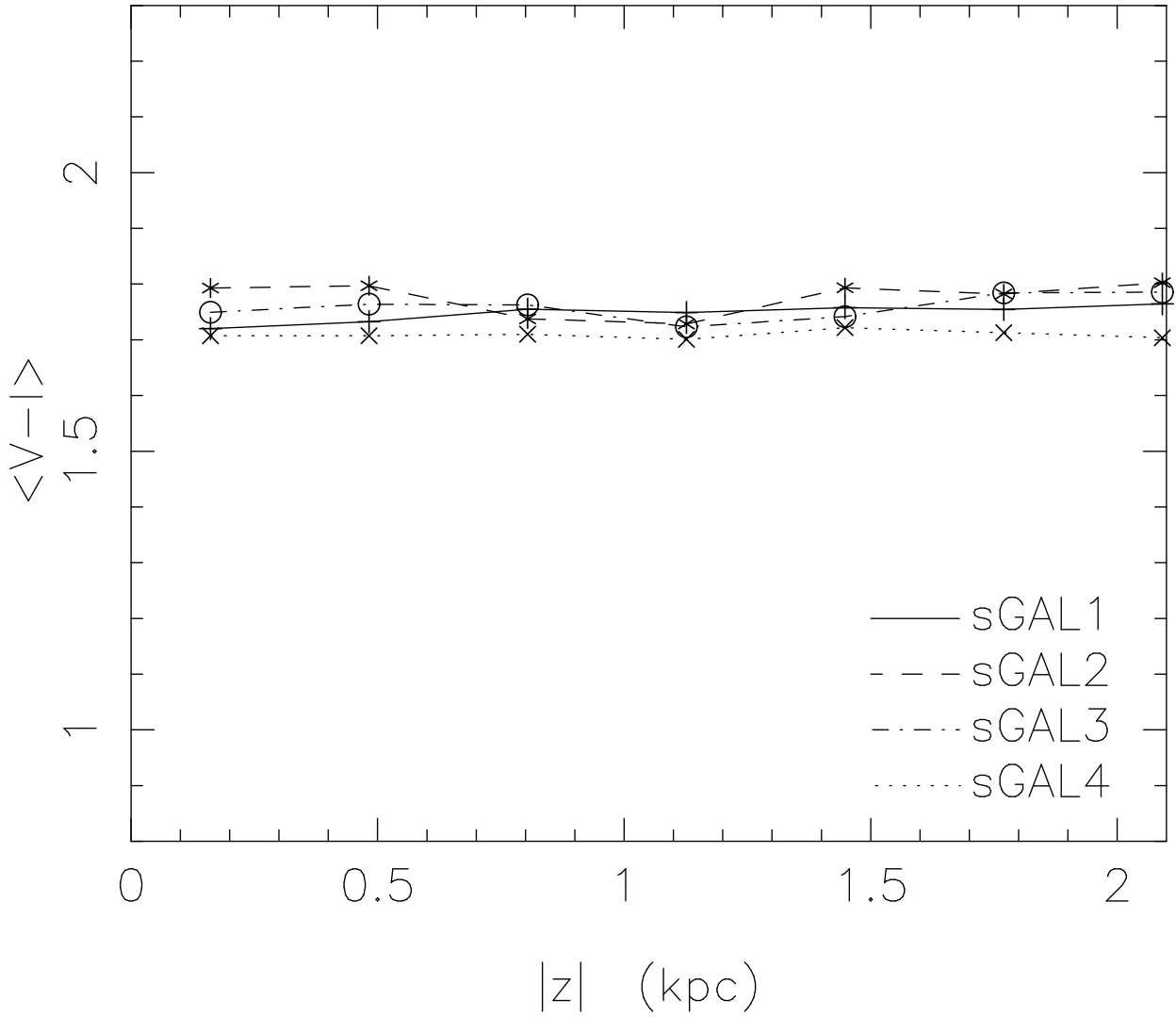}
\caption{ Mean $V-I$ color as a function of $|$Z$|$ for the 
giant branch stars (RGB/AGB) of the thick disk stars in
sGALS. The same symbols as Figure~\ref{sigW} are employed.
}
\label{VI}
\end{figure}

\begin{figure}
\epsfxsize=140.mm \plotone{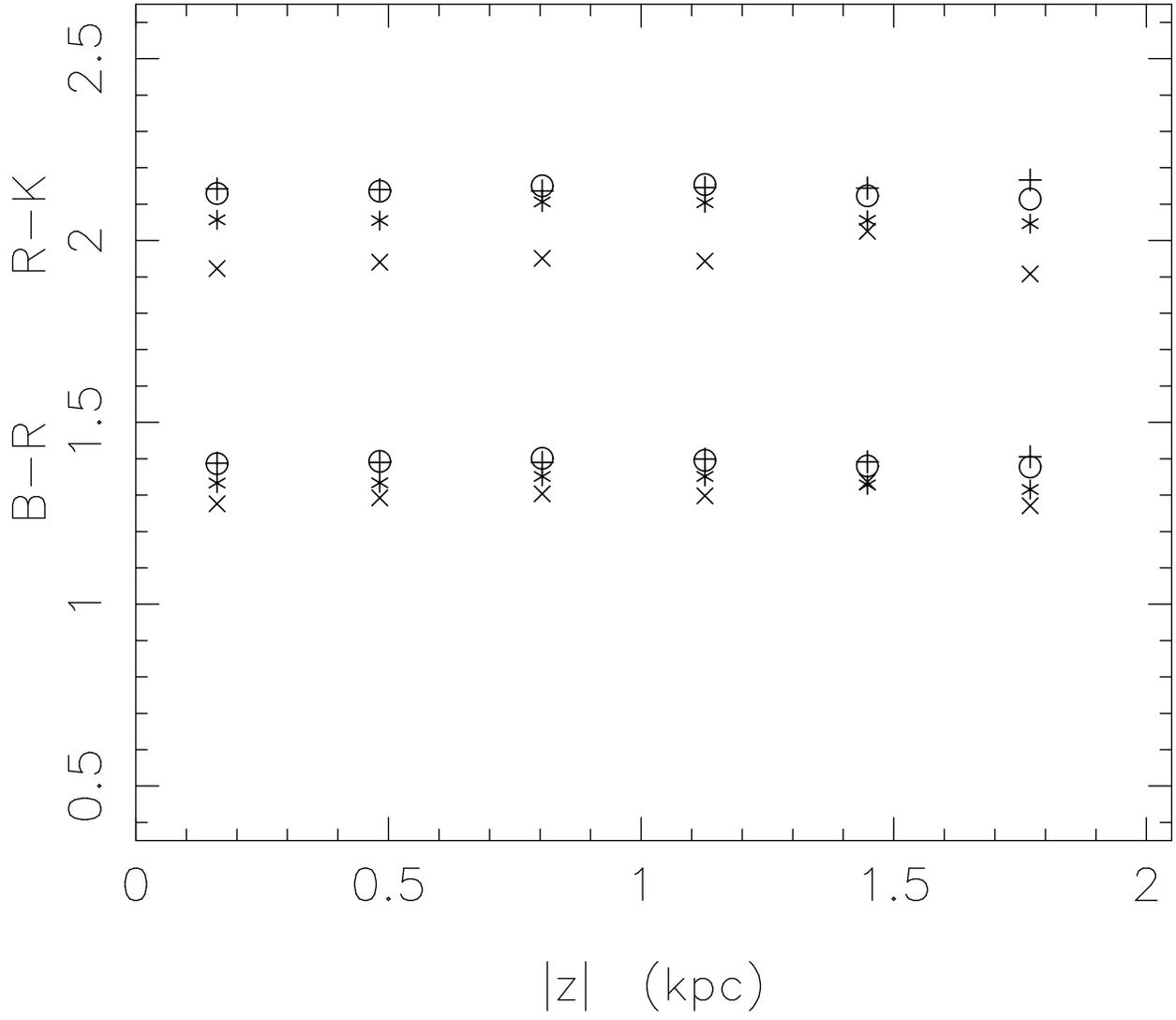}
\caption{ Mean $B$$-$$R$ and $R$$-$$K$ colors as a function of $|$Z$|$ for the
thick disk stars of sGALS. Symbols used in Figure~\ref{sigW} are employed.
}
\label{BRRK}
\end{figure}

\begin{figure}
\epsfxsize=140.mm \plotone{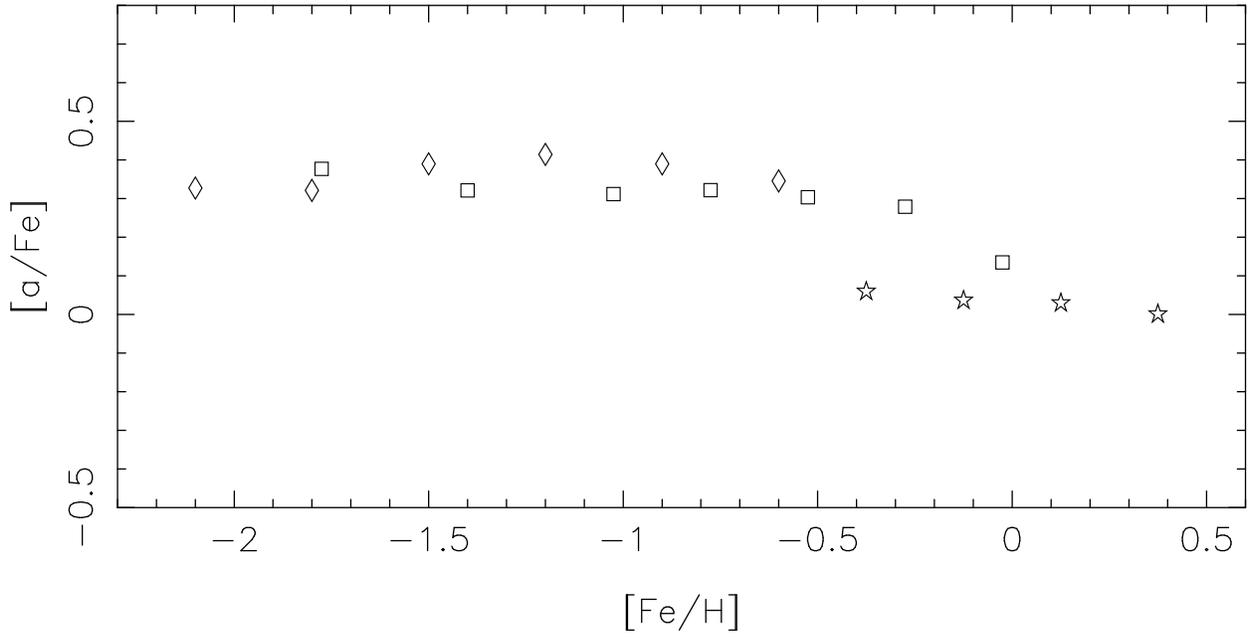}
\caption{[$\alpha$/Fe] versus [Fe/H] for solar neighborhood stars of
sGAL1. Thin disk, thick disk, and halo stars are
represented by star, square, and diamond symbols, respectively. 
}
\label{alpha1}.
\end{figure}

\begin{figure}
\epsfxsize=140.mm \plotone{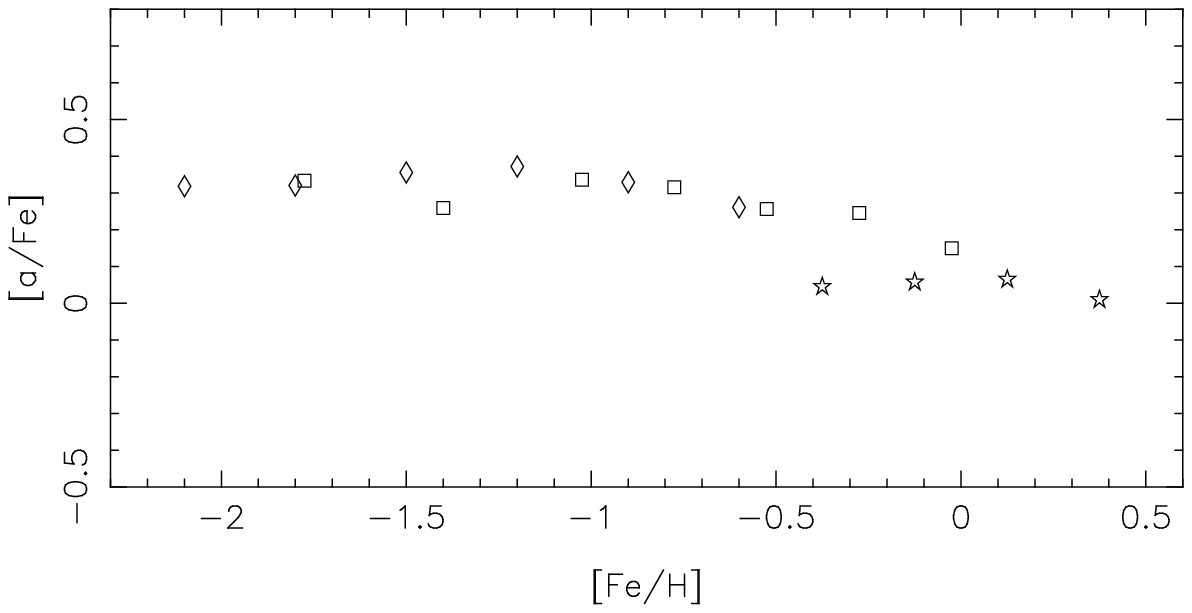}
\caption{Same as Figure~\ref{alpha1}, but for sGAL2.}
\label{alpha2}
\end{figure}

\begin{figure}
\epsfxsize=140.mm \plotone{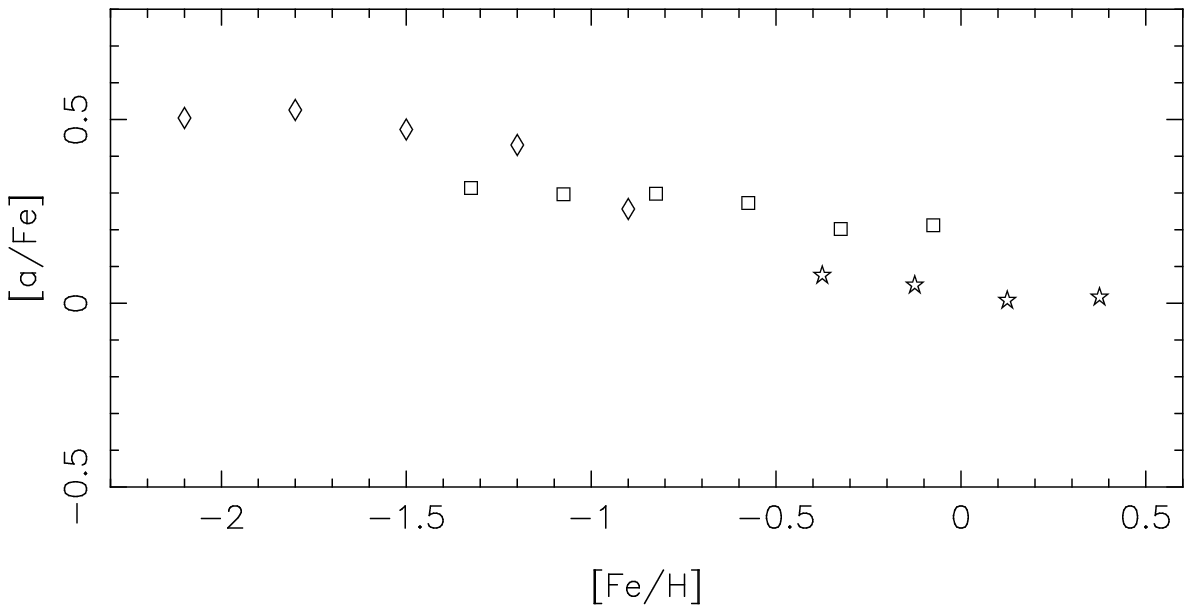}
\caption{Same as for Figure~\ref{alpha1}, but for sGAL3.}
\label{alpha3}
\end{figure}

\begin{figure}
\epsfxsize=140.mm \plotone{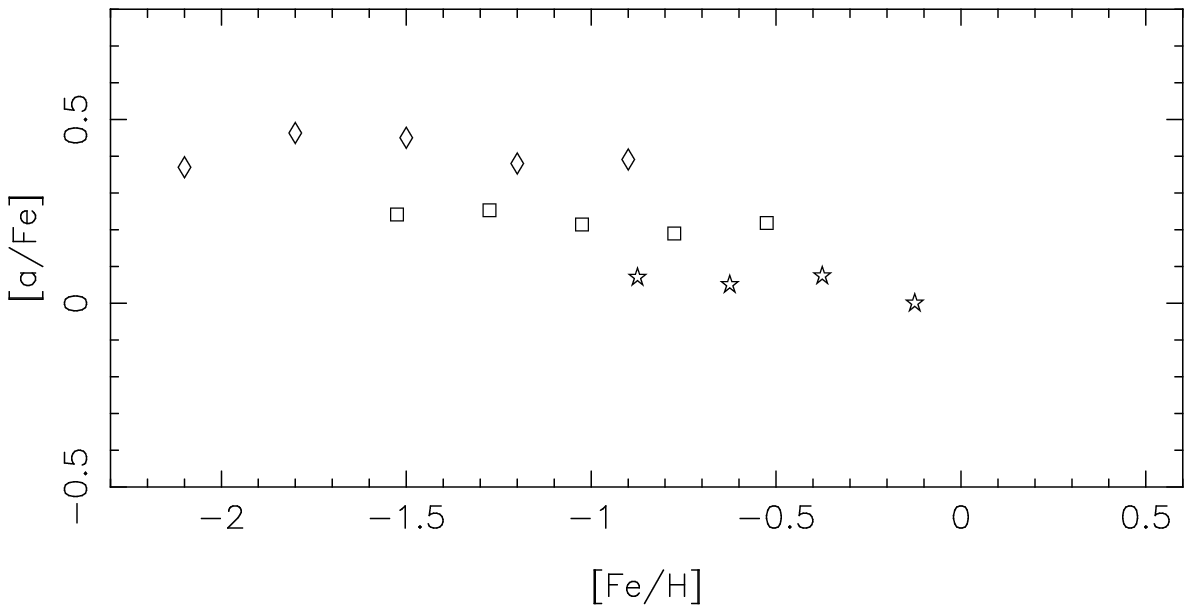}
\caption{Same as for Figure~\ref{alpha1}, but for  sGAL4.}
\label{alpha4}
\end{figure}

\begin{figure}
\epsfxsize=170.mm \plotone{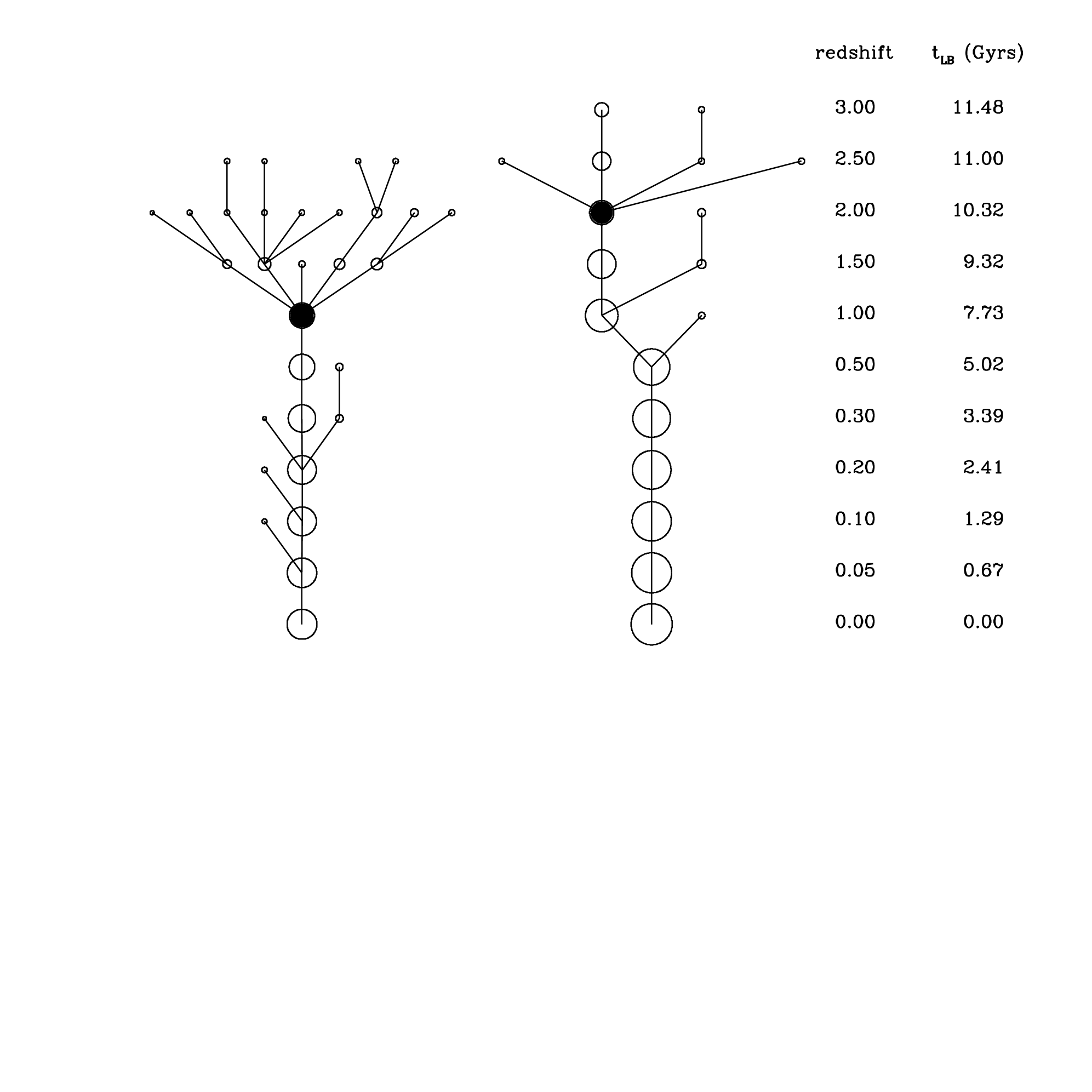}
\caption{Merger trees for two of our 58 simulated halos that are 
Milky Way analogs. The tree on the left (right) is referred to as halo 1 (2) in the text. Within each tree, the area of each circle is
proportional to the mass of the halo, and the black circle
identifies the product of the greatest merger epoch. Redshifts and lookback times are indicated on the right side.}
\label{trees}
\end{figure}

\begin{figure}
\epsfxsize=170.mm \plotone{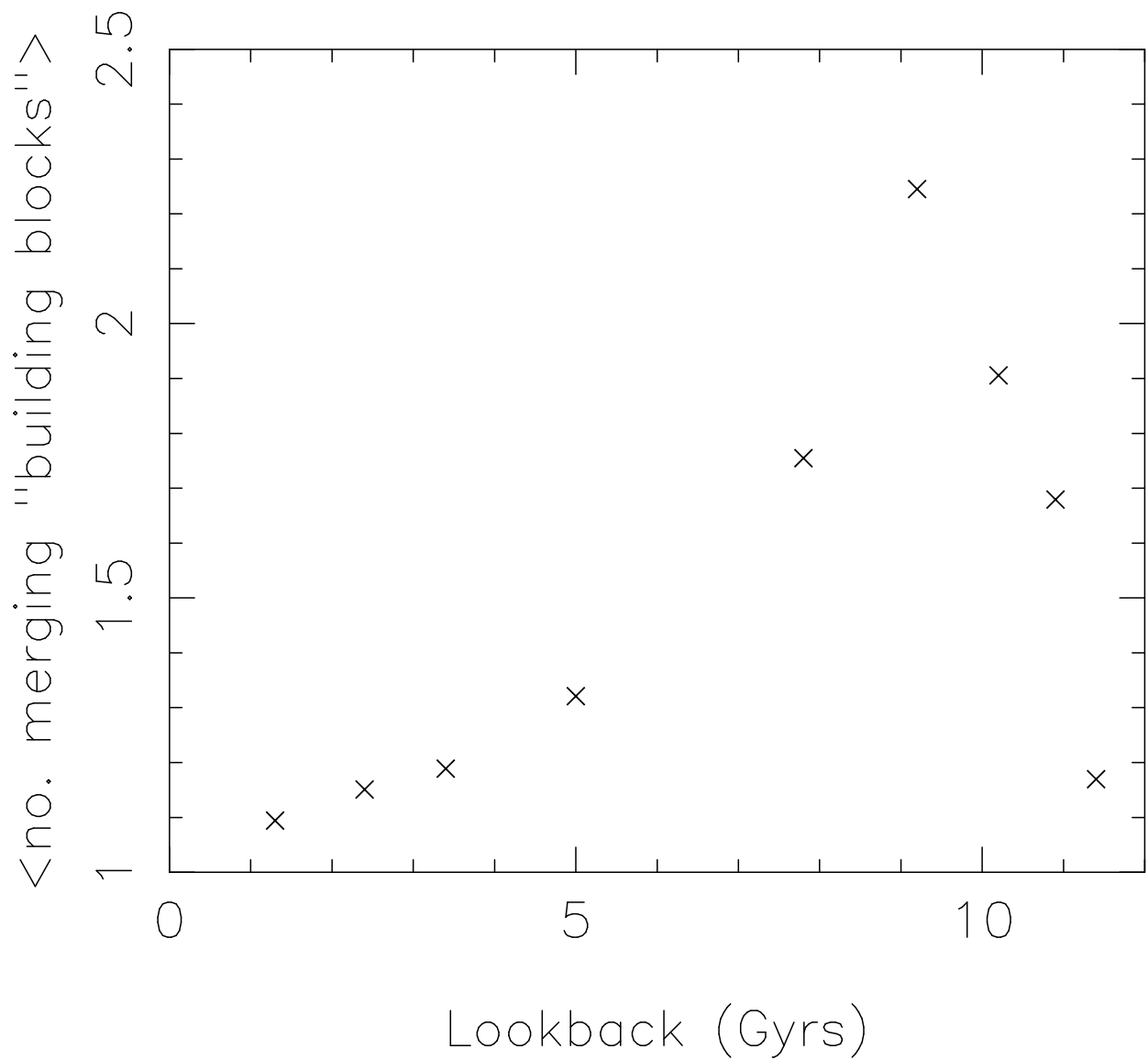}
\caption{Average number of ``building blocks'' versus lookback time for our 58 Milky Way sized dark matter halos. In order to qualify, the ``building block'' must be $\ge10^{10}\rm{M}_\odot$, add at least $4\%$ to the mass of the halo, and be merged to the largest (central) halo by the next timestep. Thus, this plot summarizes the history of ``significant merging''  in the build up of  Milky Way sized halos.}
\label{buildingblocks}
\end{figure}

\end{document}